\documentstyle[12pt]{article}

\def\lra{\longleftrightarrow}
\def\anti#1{\bar{#1}}
\def\L{{\cal L}}
\def\eps{\varepsilon}
\def\be{\begin{equation}}
\def\ee{\end{equation}}

\begin{document}

\author{N.G.Marchuk}
\title{A model of composite structure of quarks and leptons}
\maketitle
\vskip 1cm

Steklov Mathematical Institute, Gubkina st.8, Moscow 119991, Russia

nmarchuk@mi.ras.ru,   www.orc.ru/\~{}nmarchuk
\vskip 1cm


\begin{abstract}
In the model every quark or lepton is identified with a quartet of four
"more elementary" particles. One particle in a quartet is a massive spin-0 boson and
other three particles are massless spin-1/2 fermions.
\end{abstract}

\maketitle

In the Standard Model leptons and quarks of three generations
\begin{equation}
\pmatrix{ \nu_e & u \cr e^- & d}, \quad
\pmatrix{ \nu_\mu & c \cr \mu^- & s}, \quad
\pmatrix{ \nu_\tau & t \cr \tau^- & b}.\label{ff}
\end{equation}
are considered as fundamental (noncomposite) particles. In our model every quark or
lepton is identified with a quartet of four
"more elementary" particles. One particle in a quartet is a massive spin-0 boson and
other three particles are massless spin-1/2 fermions. Two iterations of the
model \cite{Marchuk1,Marchuk2} was proposed as an attempt to improve Harari-Shupe model
\cite{Shupe}-\cite{Harari3}. Our model has some common features with the
Terazawa-Chikashige-Akama model \cite{Terazawa1,Terazawa2}.
\bigskip

Let us assume the existence of ten elementary particles, which we divide in five
groups
$$
\beta^r,\beta^y,\beta^b; \quad\lambda;\quad\eps^u,\eps^d;\quad\delta;\quad
\Delta^1,\Delta^2,\Delta^3.
$$
Following \cite{Marchuk1,Marchuk2}, we call them {\em inds}.
We associate letters $\beta,\lambda,\eps$ in designations of inds with the words
"barion","lepton","electroweak" respectively.
Superscripts $r,y,b$ indicate (QCD) colors of inds  $\beta^r,\beta^y,\beta^b$;
superscripts $u,d$ show the properties upness and downness of inds
$\eps^u,\eps^d$ that associate with $u,d$ quarks and with neutrino
and electron; superscripts 1,2,3 of inds $\Delta^1,\Delta^2,\Delta^3$
indicate a generation of a particle. The quantum numbers of inds
are collected in Table 1.
\medskip

Table 1.
$$
\matrix{
\hbox{Ind}&\hbox{Q}&\hbox{B}&\hbox{L}&\hbox{color}&\hbox{spin}&\hbox{mass}&\hbox{suit} \cr
\beta & 0 & 1/3 & 0 & r,y,b & 1/2 & 0 & \diamondsuit\cr
\lambda & -2/3 & 0 & 1 & 0 & 1/2 & 0 & \diamondsuit\cr
\eps^u & 2/3 & 0 & 0 & 0 & 1/2 & 0 & \heartsuit\cr
\eps^d & -1/3 & 0 & 0 & 0 & 1/2 & 0 & \heartsuit\cr
\delta & 0 & 0 & 0 & 0 & 1/2 & 0 & \spadesuit\cr
\Delta & 0 & 0 & 0 & 0 & 0 & m_1,m_2,m_3 & \clubsuit
}
$$
\medskip
where  $Q$ - electric charge in
units of proton charge; $B$ - barion number; $L$ - lepton number;
color - QCD color charge. I don't know the real values of masses $m_1,m_2,m_3$, but
I suppose that $m_1<m_2<m_3$.  {\em A suit} is a new quantum number, which select admissible
quartets of inds from all 715
quartets of inds. Namely quartets with full set of suits
$(\diamondsuit\heartsuit\spadesuit\clubsuit)$ we identify with fundamental fermions
(\ref{ff}): the quartets $\beta\eps\delta\Delta$ (18 pieces) we identify with quarks
and the quartets $\lambda\eps\delta\Delta$ (6 pieces) we identify with leptons.
\medskip

Table 2.
$$
\matrix{
\hbox{ff}&\hbox{quartet}&\hbox{Q}&\hbox{B}&\hbox{L}&\hbox{color} \cr
\nu_e & \lambda\eps^u\delta\Delta^1 & 0 & 0 & 1 & 0 \cr
e^- & \lambda\eps^d\delta\Delta^1 & -1 & 0 & 1 & 0 \cr
u & \beta\eps^u\delta\Delta^1 & 2/3 & 1/3 & 0 & r,y,b \cr
d & \beta\eps^d\delta\Delta^1 & -1/3 & 1/3 & 0 & r,y,b
}
$$
\medskip

We get tables for fundamental fermions of second and third generations by replacing
ind $\Delta^1$ by $\Delta^2$ and $\Delta^3$ respectively.
Fermion inds $\beta,\lambda,\eps,\delta$ are massless and we may consider left-handed
and right-handed inds as different particles. We suppose that left-handed inds
$\eps^u_L,\eps^d_L$ have the weak isospin $I^w=1/2$. All other inds have $I^w=0$.
In the following table an electric charge $Q$ connected with a weak hypercharge
$Y^w$ and with a third component of weak isospin by the Gell-Mann--Nishijima formula
$Q=I^w_3+Y^w/2$
\medskip

Table 3.
$$
\matrix{
\hbox{ind}& Q& Y^w & I^w & I^w_3\cr
\eps^u_L & 2/3 & 1/3 & 1/2 & 1/2\cr
\eps^u_R & 2/3 & 4/3 & 0 & 0\cr
\eps^d_L & -1/3 & 1/3 & 1/2 & -1/2\cr
\eps^d_R & -1/3 & -2/3 & 0 & 0
}
$$
\medskip

Using principles of Standard Model, we can write down the fundamental Lagrangian
${\cal L}_{\rm inds}$, which describes a
dynamics of particles on inds level
\begin{eqnarray}
\L_{inds}&=&\{\anti\Psi i \gamma_\mu(\partial_\mu-
{i g_3\over2}\lambda^l C^l_\mu)\Psi -{1\over4}C^l_{\mu\nu}C^l_{\mu\nu}\}\cr
&+&\{
\anti \Phi i\gamma_\mu(\partial_\mu+(\frac{4}{3}){i g_1\over2} B_\mu)\Phi \cr
&+&\anti L i \gamma_\mu (\partial_\mu-(\frac{1}{3}){i g_1\over 2}B_\mu-
{i g_2\over2}\tau^k W^k_\mu)L \cr
&+&\anti R^u i\gamma_\mu(\partial_\mu-(\frac{4}{3}){i g_1\over2} B_\mu)R^u \cr
&+&\anti R^d i\gamma_\mu(\partial_\mu+(\frac{2}{3}){i g_1\over2} B_\mu)R^d \cr
&-&{1\over4}B_{\mu\nu}B_{\mu\nu}-{1\over4}W^k_{\mu\nu}W^k_{\mu\nu}\cr
&+&{|D_\mu \phi|}^2-{1\over2}\alpha^2(|\phi|^2-{1\over2}\eta^2)^2\}\cr
&+&\{\anti Y i\gamma_\mu\partial_\mu Y\}\cr
&+&\{\sum_{k=1}^3(\frac{1}{2}\partial_\mu\omega_k
\partial^\mu\omega_k-\frac{m_k{}^2}{2}\omega_k{}^2)\}
\end{eqnarray}
where we sum over $\mu,\nu=0,1,2,3$; $k=1,2,3$;
$l=1,\ldots,8$; $\tau^k$ -- Pauli matrices, $\lambda^l$ -- Gell-Mann
matrices.
$\phi=\pmatrix{\phi_+ \cr \phi_0}$ is a complex dublet of scalar fields
with
$Y^w=1$. The covariant derivatives have the form
$$
D_\mu=\partial_\mu - {i g_1\over2}B_\mu - {i g_2\over2}\tau^k W^k_\mu;
$$
$\alpha,\eta$ are real constants. After the gauge transformation the
dublet
$\phi$ has the form
$$
\phi={1\over{\sqrt2}}\pmatrix{0\cr \eta + \chi(x)},
$$
where constant $\eta$ is a vacuum average and $\chi$ is a real scalar
field. $\Psi,\Phi$, $L,R^u,R^d,Y$ are spinor wave functions of the
following multiplets of inds:
$$
\Psi=\pmatrix{\beta^r\cr \beta^y\cr \beta^b},\
\Phi=\pmatrix{\lambda},\
L=\pmatrix{\eps^u_L\cr \eps^d_L},\
R^u=\pmatrix{\eps^u_R}, \
R^d=\pmatrix{\eps^d_R},\
Y=(\delta)
$$
and
$\omega_k=(\Delta^k)$ are scalar wave functions of inds $\Delta^k$.

$\Psi$ is an SU(3)-triplet,
$L$ is an ${\rm U}(1)\!\times\!{\rm SU}(2)$-doublet,
$R^u,R^d,\Phi$ are U(1)-singlets, and
$Y$ is a singlet without gauge symmetry.
Wave functions of fermion inds satisfy Dirac equations and wave functions of boson
inds satisfy Klein-Gordon equations.
$\anti\Psi, \anti\Phi,\anti L,\anti R^u,\anti R^d, \anti Y$ are
spinorial conjugated wave functions.
$g_1,g_2$ are the electroweak coupling constants.
$g_3$ is the strong coupling constant,
\begin{eqnarray*}
        B_{\mu\nu} &=& \partial_\mu B_\nu - \partial_\nu B_\mu,\cr
        W_{\mu\nu} & =& \partial_\mu W_\nu - \partial_\nu W_\mu -
           i g_2 [W_\mu, W_\nu] = W_{\mu\nu}^k {\tau^k\over2},
           \quad W_\mu=W_\mu^k{\tau^k\over2} \cr
        C_{\mu\nu} & =& \partial_\mu C_\nu - \partial_\nu C_\mu -
           i g_3 [C_\mu, C_\nu] = C_{\mu\nu}^l {\lambda^l\over2},
           \quad C_\mu=C_\mu^l{\lambda^l\over2}
\end{eqnarray*}
Values
$C_\mu^l$ define vector fields of gluons. The vector field
$A_\mu$ of photon $\gamma$ and vector fields
 $Z_\mu^0,W_\mu^\pm$ of $Z^0, W^\pm$ bosons calculated as
\begin{eqnarray*}
A_\mu &=&{1\over{\sqrt{ {g_1}^2 + {g_2}^2 }}}(g_2 B_\mu + g_1 W_\mu^3)\cr
Z_\mu^0 &=&{1\over{\sqrt{ {g_1}^2 + {g_2}^2 }}}(- g_1 B_\mu + g_2 W_\mu^3)\cr
W_\mu^\pm &=& {1\over\sqrt{2}}(W_\mu^1 \mp i W_\mu^2)
\end{eqnarray*}
The term $|D_\mu \phi|^2$ gives mass terms for the intermediate bosons
(indices $\mu$ not written)
$$
{\eta^2\over8}(g_2 W^3-g_1 B)^2+{\eta^2 {g_2}^2\over2}W^- W^+
$$
and masses
$$
m_Z={\eta \sqrt{{g_1}^2+{g_2}^2}\over2}, \quad m_W={\eta g_2\over2},
\quad {m_W\over{m_Z}}=\cos{\theta_W}.
$$
Usual formulas connect constants
$g_1,g_2$ with the electromagnetic coupling constant
$e$ and with the Weinberg angle $\theta_W$
$$
e = {g_1 g_2\over \sqrt{ {g_1}^2 + {g_2}^2} } = g_2 \sin{\theta_W},
\qquad \tan{\theta_W} = {g_1\over g_2}.
$$
A mass of scalar field
$\chi$ (Higgs particle $H$) has the form
$m_H=\lambda\eta$. A connection between the vacuum average $\eta$ and
the Fermi
constant
$G$ is given by the formula $\eta=(\sqrt2 G)^{-1/2}$.
\medskip

Note that in the present model there is a rather attractive
possibility to use a technicolor mass generation mechanism for
$Z^0,W^\pm$ bosons.

The Lagrangian $\L_{inds}$ is written as a sum of four independent
terms in curly brackets. The first term for inds
$\beta$ discribes strong interactions. The second term for inds
$\eps$ and $\lambda$ discribes  electroweak interactions.
So, the Lagrangian
$\L_{inds}$ discribes strong and electroweak
interactions simultaneously and, in contrast to the Standard
Model,  the present model unify strong
and electroweak interactions.
\medskip

According to the principles of Standard Model, we consider fundamental bosons --
$\gamma,Z^0,W^\pm,g_1,\ldots,g_8$ ($g_k$ -gluons of QCD) as gauge fields of
${\cal L}_{\rm inds}$ relevant to
${\rm U}(1),{\rm U}(1)\!\times\!{\rm SU(2)}_L,{\rm SU}(3)$ gauge symmetries.

From the other side we may consider fundamental bosons --
$\gamma,Z^0,W^\pm$, $g_1,\ldots,g_8$ as composite particles that consist of
ind-antiind pairs (superpositions of pairs) with $\diamondsuit\bar\diamondsuit$ or
$\heartsuit\bar\heartsuit$ suits. Namely \footnote{Formulae for $\gamma$
and $Z^0$ are not indisputable.}
\begin{eqnarray*}
W^+&=&\eps^u\bar\eps^d,\\
W^-&=&\bar\eps^u\eps^d,\\
Z^0&=&\frac{1}{\sqrt{2}}(\eps^u\bar\eps^u-\eps^d\bar\eps^d){\rm cos}\theta_W+
\lambda\bar\lambda{\rm sin}\theta_W,\\
\gamma&=&-\frac{1}{\sqrt{2}}(\eps^u\bar\eps^u-\eps^d\bar\eps^d){\rm sin}\theta_W+
\lambda\bar\lambda{\rm cos}\theta_W,\\
g^{y\bar{b}}&=&\beta^y\bar\beta^b, \ldots,
\end{eqnarray*}
where $\theta_W$ is the Weinberg angle.

Strong and electroweak interactions of particles with the participation
of fundamental bosons
can be interpreted as an exchange of constituent parts (inds) of
interacting particles. Fundamental bosons are particles that accomplish such an
exchange. For example, let us consider several
reactions and their interpretations on inds level
\begin{eqnarray*}
u^r+g^{y\anti r}\lra u^y &\equiv&
\beta^r\,\eps^u\,\delta\Delta^1+\beta^y\,\anti\beta^r\lra\beta^y\,\eps^u\,\delta\Delta^1\\
g^{r\anti b}+g^{b\anti y}\lra g^{r\anti y} &\equiv&
\beta^r\,\anti\beta^b+\beta^b\,\anti\beta^y\lra\beta^r\,\anti\beta^y\\
e^-+W^+\lra\nu_e &\equiv&
\lambda\,\eps^d\,\delta\Delta^1+\eps^u\,\anti\eps^d\lra\lambda\,\eps^u\,\delta\Delta^1\\
d+W^+\lra u &\equiv&
\beta\,\eps^d\,\delta\Delta^1+\eps^u\,\anti\eps^d\lra\beta\,\eps^u\,\delta\Delta^1
\end{eqnarray*}

This interpretation is similar to the interpretation on quarks level of
pion exchange between nucleons
$$
p+\pi^-\lra n\quad\equiv\quad uud+\anti u d\lra udd.
$$

There is an evident analogy between the present model,
which consider quarks and leptons as quartets of inds, and quantum
chromodynamics, which consider barions as trios of quarks. This analogy leads us to the
assumption that all inds have one more quantum number with four
possible values, which we call {\em 4-color}.
Suppose that quarks and leptons are 4-color singlets
(states antisymmetric in 4-color). Suppose also an existence of
4-color gluons, which are quanta of 4-color
interaction that confines quartets of inds in quarks and leptons.
It seems natural to
describe a 4-color interaction as a gauge field theory with ${\rm SU}(4)$
symmetry (15 generators)
that corresponds to 4-color.

So in the model we use gauge groups
${\rm U}(1)$,${\rm U}(1)\!\times\!{\rm SU}(2)$,${\rm SU}(3)$,${\rm SU}(4)$,
which are subgroups of the group ${\rm U}(4)$. According to
\cite{Marchuk3,Marchuk4,Marchuk5}, for these gauge groups Dirac-type tensor equations can be used
instead of Dirac equations.

There are many things to do for development of the model. In partiqular,
it is important to understand a dynamical mechanism that confines three
massless fermions and one massive boson in a particle with a dimension
less than $10^{-17}cm$. Some arguments of G.'t Hooft \cite{Hooft} may be
usefull.


\begin{thebibliography}{99}
\bibitem{Marchuk1} N.G.Marchuk, Doklady of Russian Academy of Science 344, 1 (1995).

\bibitem{Marchuk2} N.G.Marchuk, A model of the composite structure of quarks and leptons
with SU(4) gauge symmetry, hep-ph/9801382.

\bibitem{Marchuk3} N.G.Marchuk, Nuovo Cimento, 117B, N.1, (2002), p.95.

\bibitem{Marchuk4} N.G.Marchuk, Dirac-type tensor equations on parallelisable
manifolds, Nuovo Cimento, 117B, 7 (2002).

\bibitem{Marchuk5} N.G.Marchuk, A concept of Dirac-type tensor equations,
        math-ph/0212006.

\bibitem{Shupe} M.A.Shupe, Phys. Lett. 86B (1979) 87.

\bibitem{Harari1} H.Harari, Phys. Lett. 86B (1979) 83.

\bibitem{Harari2} H.Harari and N.Seiberg, Phys. Lett. 98B (1981) 269.

\bibitem{Harari3} H.Harari and N.Seiberg, Phys. Lett. 100B (1981) 41.

\bibitem{Terazawa1} H.Terazawa, Y.Chikashige and K.Akama, Phys. Rev. D15 (1977) 480.

\bibitem{Terazawa2} H.Terazawa, Phys. Rev. D22 (1980) 184.

\bibitem{Hooft} G.'t Hooft, in "Recent Developments in Gauge Theories",
Plenum Press (1980).

\end{thebibliography}
\end{document}